\documentstyle[12pt]{article}

\csname @addtoreset\endcsname{equation}{section}

\def\bseq{\begin{subequation}}  % = 1a 1b
\def\eseq{\end{subequation}}
\def\bsea{\begin{subeqnarray}}  % = 1.1a 1.1b
\def\esea{\end{subeqnarray}}
\newcommand{\bbox}{\lower.2ex\hbox{$\Box$}}

\newcommand{\beq}{\begin{equation}}
\newcommand{\eeq}{\end{equation}}
\newcommand{\bea}{\begin{eqnarray}}
\newcommand{\eea}{\end{eqnarray}}
\newcommand{\ena}{\end{eqnarray}}

\renewcommand{\a}{\alpha}
\renewcommand{\b}{\beta}

\renewcommand{\d}{\delta}

\newcommand{\pa}{\partial}
\newcommand{\g}{\gamma}

\newcommand{\e}{\epsilon}

\renewcommand{\L}{\Lambda}
\newcommand{\m}{\mu}

\newcommand{\n}{\nu}

\newcommand{\p}{\pi}

\newcommand{\s}{\sigma}

\begin{document}

\begin{titlepage}
\begin{flushright} IFUM--632--FT\\
%\\ hep-th/98yyxxx
\end{flushright}
\vfill
\begin{center}
{\LARGE\bf Gravitational corrections\\
\vskip 3.mm
 to $N=2$ supersymmetric lagrangians}\\
\vskip 27.mm  
{\large\bf  Sergio Cacciatori and  Daniela Zanon  } \\
\vskip 5.mm
{\small
Dipartimento di Fisica dell'Universit\`a di Milano and\\
INFN, Sezione di Milano, via Celoria 16,
I-20133 Milano, Italy}\\
\end{center}
\vfill

\begin{center}
{\bf ABSTRACT}
\end{center}
\begin{quote}
In the framework of special K\"{a}hler geometry we consider the 
supergravity-matter system which emerges on a $K3$-fibered Calabi-Yau 
manifold. By applying the rigid limit procedure in the vicinity of a
conifold singularity we compute the K\"{a}hler potential of the scalars
and the kinetic matrix of the vectors to first order in the gravitational
couplings.

\vfill     
\vskip 5.mm
 \hrule width 5.cm
\vskip 2.mm
{\small
\noindent e-mail: sergio.cacciatori@mi.infn.it\\
\noindent e-mail: daniela.zanon@mi.infn.it}
\end{quote}
\begin{flushleft}
September 1998
\end{flushleft}
\end{titlepage}

%\section{Introduction}
Type II string theory compactified on a Calabi-Yau threefold leads to
$N=2$ supersymmetric theories in four dimensions. 
The geometric structure
underlying the four-dimensional models is special K\"{a}hler geometry 
\cite{Sugra, rigid, strominger} in its
local or rigid version depending on whether the matter fields are coupled or  
decoupled to supergravity, respectively. In type IIB compactifications
the classical geometry of the Calabi-Yau complex structure moduli space
exhibits all the defining
properties of a special K\"{a}hler manifold for vector multiplets coupled to
$N=2$ supergravity \cite{seiberg, candelas}. The classical geometry does not receive 
quantum corrections and consequently exact results for the 
corresponding low-energy effective action can be obtained. Indeed,
starting from a specific special K\"{a}hler manifold, once we know its symplectic
vectors, the full supergravity-matter action can be constructed, i.e. the
K\"{a}hler potential of the scalar fields and the kinetic matrix of the vectors
can be determined. In the Calabi-Yau moduli space language this amounts
to the computation of the period vectors. In principle given the local 
version of the theory it should be straightforward to perfom the rigid limit
and obtain the action for scalars and vectors in flat space. In practice it 
turns out that it is not a priori obvious how to decouple gravitons, 
gravitini and graviphotons in a supersymmetric invariant way.

In ref. \cite{Lerche, KLM, L} the rigid limit procedure has been applied 
directly in the moduli
space of the Calabi-Yau threefold with an expansion in the vicinity
of singular points. In particular in \cite{7authors} a detailed study has 
been presented for Calabi-Yau surfaces which are $K3$ fibrations. There the
reduction from local to rigid special K\"{a}hler geometry has been
illustrated on specific examples. Exact expressions for all the periods 
have been obtained and the corresponding K\"{a}hler potential has been
computed explicitly in the rigid limit.

In this letter we progress further in the construction of supergravity-matter
actions, namely we consider one of the Calabi-Yau manifolds
studied in \cite{7authors} and  by applying the rigid limit expansion in 
the neighbourhood of a
conifold singularity we evaluate the K\"{a}hler potential of the scalars
and the kinetic matrix of the vectors to {\em first order} in the gravitational
couplings.

\vspace{0.6cm}

The Calabi-Yau surface we consider corresponds to the zero-loci of the
following quasi-homogeneous polynomial
\beq
W=\frac{b_1}{8} x_1^8 +\frac{b_2}{8} x_2^8+ \frac{b_3}{4} x_3^4
+\frac{b_4}{4} x_4^4+\frac{b_5}{4} x_5^4-\psi_0 x_1 x_2 x_3 x_4 x_5
-\frac{1}{4}\psi_s(x_1x_2)^4
\label{poly}
\eeq
in the weighted projective space $CP_8(1,1,2,2,2)$ with global 
identifications
\beq
x_j\sim {\rm{exp}}(n_j \frac{2\p i}{8}) x_j
\label{globalident}
\eeq
with
\bea
&&(n_1,n_2,n_3,n_4,n_5)= m_0(1,1,2,2,2)+m_1(1,-1,0,0,0)\nonumber\\
&&~~~~~~~~~~~~~~~~~~~~~~~+m_2(0,0,2,-2,0)+m_3(0,0,2,0,-2)
\eea
where $m_0$, $m_1$, $m_2$, $m_3$ are integers.
There are only two independent moduli, since in (\ref{poly}) rescalings of the
$x_i$ variables induce five gauge invariances. It is useful to perform a 
partial gauge-fixing by setting
\beq
b_1=b_2=B \qquad\qquad\qquad b_3=b_4=b_5=1
\label{gaugefixing}
\eeq
Introducing new variables $x_0$ and $\zeta$ with
\beq
x_1= \zeta^{\frac{1}{8}}\sqrt{x_0} \qquad\qquad 
x_2= \zeta^{-\frac{1}{8}}\sqrt{x_0}
\label{defzeta}
\eeq
and defining
\beq
B'=\frac{1}{2}(B\zeta+\frac{B}{\zeta}-2\psi_s)
\label{defB'}
\eeq
one explicitly exhibits the $K3$ fibration of the Calabi-Yau manifold:
for fixed $\zeta$ the fiber is given by the zero-loci of the polynomial
\beq 
W=\frac{1}{4}\left( B'x_0^4+x_3^4+x_4^4+x_5^4\right)-\psi_0 x_0x_3x_4x_5
\label{K3fiber}
\eeq
in the weighted projective space $CP_4(1,1,1,1)$. On the $K3$ manifold
the identifications in (\ref{globalident}) become  
\beq
x_j\sim {\rm {exp}}(n_j \frac{2\p i}{4}) x_j
\label{globalidentK3}
\eeq
with
\beq
(n_0,n_3,n_4,n_5)= m_0(1,1,1,1)+m_2(0,1,-1,0)+m_3(0,1,0,-1)
\eeq

The analysis of the singularities in the moduli space of the $K3$ and the
Calabi-Yau manifolds has been presented in great detail in 
ref. \cite{7authors}. Here we briefly recall the main results: 
singularities occur
when $W=0$ and $dW=0$. Thus the $K3$ fiber becomes
singular at $B'=0$ and at $\psi_0^4=B'$. Correspondingly in the $\zeta$ 
plane, due to the symmetry $\zeta\rightarrow 1/\zeta$, one finds the points
\beq
e^{\pm}_0= \frac{1}{B}\left(\psi_s\pm \sqrt{\psi_s^2-B^2}\right) 
\qquad\qquad\qquad
e^{\pm}_1=\frac{1}{B}\left(\psi_s+\psi_0^4\pm 
\sqrt{(\psi_s+\psi_0^4)^2-B^2}\right)
\label{singularities}
\eeq
For the Calabi-Yau manifold to become singular one has to impose the 
additional condition $\pa_{\zeta}W=0$ which leads to the vanishing of the
discriminant
\beq
\Delta_{CY}=B^6(e^+_0-e^-_0)^2(e^+_1-e^-_1)^2
\eeq
The location of the 
singularity which is of interest for the rigid limit is at $B=0$. Therefore,
choosing as independent moduli
\beq
B \qquad\qquad{\rm{and}}\qquad\qquad \tilde{u}=\frac{\psi_s+\psi_0^4}{B}
\eeq
the neighbourhood of the singular point in moduli space is parametrized 
by $B=2\e$, and $\psi_s+\psi_0^4=2\e \tilde{u}$, keeping $\tilde{u}$ fixed 
and expanding in powers of $\e$.

The fibration structure exhibited in (\ref{K3fiber}) allows to express 
the volume form and the highest 
degree holomorphic form of the Calabi-Yau manifold in terms of the 
corresponding ones on the fiber
\bea
&&\omega_{CY}= \frac{1}{4} \omega_{K3}\frac{d\zeta}{\zeta}
\qquad\qquad\qquad \Omega^{(3,0)}=\Omega^{(2,0)}\frac{d\zeta}{2\p i \zeta}
\nonumber\\
&&~~~~~~~~~~~\nonumber\\
&&\omega_{K3}=x_0dx_3dx_4dx_5-dx_0(x_3dx_4dx_5+x_4dx_5dx_3+x_5dx_3dx_4)
\eea
This factorized expression drastically simplifies the evaluation of the 
periods of the Calabi-Yau manifold \cite{7authors}: first one integrates 
over two-cycles of the $K3$ manifold thus obtaining three linearly 
independent periods, then the $\zeta$-integral is performed choosing
paths which are relevant in view of the rigid limit of the supergravity
theory.

In this way one obtains six 
Calabi-Yau periods, which are holomorphic functions of the moduli. 
The periods and their intersection matrix are identified with the
symplectic vectors and the symplectic metric, respectively, of special 
K\"{a}hler
geometry.  Finally one applies standard procedures 
\cite{specialgeom, whatisspecial}
and computes the local version of the
K\"{a}hler potential and the kinetic matrix of the vector fields. 
Indeed let us denote by
\beq
v=\left( \matrix{v_1\cr v_2\cr v_3\cr v_4\cr v_5\cr v_6\cr}\right)
\label{periodvector}
\eeq
the period vector whose components are given by the six Calabi-Yau periods
evaluated in a basis in which the corresponding intersection matrix is 
the canonical symplectic metric
\beq
q=\left(\matrix{0 &1\cr -1& 0\cr}\right)
\label{canonical}
\eeq
The symplectic inner product is defined as
\beq
<v,w>=v^T q^{-1} w
\label{innerproduct}
\eeq
As noted above the periods, given by integration of the $(3,0)$-form over
the six three-cycles, are holomorphic functions of the independent
moduli denoted collectively by $z$. In terms of 
(\ref{periodvector}) and
(\ref{canonical}) the local K\"{a}hler potential is given by
\beq
{\cal{K}}(z,\bar{z})= -{\rm {ln}} (-iv^T(z) q^{-1} \bar{v}(\bar{z}))
\label{Kahlerpot}
\eeq
With these definitions it is easy to verify that all the requirements of
local special K\"{a}hler geometry are satisfied \cite{whatisspecial}, 
in particular
\beq
<{\cal{D}}_\a v,{\cal{D}}_\b v >=0
\eeq
where the covariant derivatives are
\beq
{\cal{D}}_\a v=\pa_\a v+(\pa_\a {\cal{K}})v \qquad\qquad\qquad
{\cal{D}}_{\bar{\a}} v=\pa_{\bar{\a}}v=0
\label{covder}
\eeq
with
\beq
\pa_\a=\frac{\pa}{\pa z^\a} \qquad,\qquad \a=1,2
\eeq
We make contact with the $N=2$ field theory actions through the additional
identifications
\beq
v=\left(\matrix{\chi^I\cr f_I\cr}\right)
\eeq
and
\beq
{\cal{N}}_{IJ}=\left(\bar{{\cal{D}}}_{\bar{\a}}\bar{f}_I, f_I\right)
\left(\bar{{\cal{D}}}_{\bar{\a}}\bar{\chi}^J, \chi^J\right)^{-1}
\eeq
so that we have for the scalars
\beq
L_0= -g_{\a\bar{\b}} D_\m z^\a D^\m \bar{z}^{\bar{\b}}
\label{actionscalars}
\eeq
and for the vectors
\beq
L_1=\frac{1}{4}~ ({\rm {Im}}{\cal{N}}_{IJ}) F_{\m\n}^{I}F^{\m\n J}-\frac{i}{8}
({\rm{Re}}{\cal{N}}_{IJ})\e^{\m\n\rho\s}F_{\m\n}^IF_{\rho\s}^J
\label{actionvectors}
\eeq
In (\ref{actionscalars}) $g_{\a\bar{\b}}$ is the K\"{a}hler metric
\beq
g_{\a\bar{\b}}=\pa_\a \pa_{\bar{\b}} {\cal{K}}
\label{Kahlermetric}
\eeq
and in (\ref{actionvectors}) $F_{\m\n}^{I}$ are the field strengths of the 
vector fields.

For our specific case the main steps summarized above are exemplified in 
detail in ref. \cite{7authors}. The formulae which are relevant for our
subsequent calculations are the following: an integral basis of periods on
the $K3$ fiber is given by
\bea
&&\hat{\vartheta}'_0=-\frac{1}{4\p^2}(U_1-U_2)^2 +
\frac{i}{4\p^2}(U_1-iU_2)^2 \nonumber\\
&&\hat{\vartheta}'_1=\frac{1}{4\p^2}(U_1-U_2)^2-\frac{2i}{4\p^2}(U_1-iU_2)^2
 +\frac{1}{4\p^2}(U_1+U_2)^2 \nonumber\\
&&\hat{\vartheta}'_2=\frac{1}{4\p^2}(U_1-U_2)^2
\label{K3periods}
\eea
where $U_{1,2}$ are two linearly independent solutions of the hypergeometric
equation with parameters $\left\{ \frac{1}{8}, \frac{3}{8},1\right\}$.
In terms of the $\hat{\vartheta}'_I$ basis one constructs six Calabi-Yau
periods
\bea
&&{\cal{T}}_v= \frac{1}{2\p i} \int_C \frac{d\zeta}{\zeta} \hat{\vartheta}'_0
=\frac{1}{\p}\int_{-1}^1\frac{dw}{\sqrt{1-w^2}}\hat{\vartheta}'_0
\nonumber\\
&&{\cal{T}}_1= \frac{1}{2\p i} \int_C \frac{d\zeta}{\zeta} \hat{\vartheta}'_1
=\frac{1}{\p}\int_{-1}^1\frac{dw}{\sqrt{1-w^2}}\hat{\vartheta}'_1
\nonumber\\
&&{\cal{T}}_2= \frac{1}{2\p i} \int_C \frac{d\zeta}{\zeta}\hat{ \vartheta}'_2
=\frac{1}{\p}\int_{-1}^1\frac{dw}{\sqrt{1-w^2}}\hat{\vartheta}'_2
\nonumber\\
&&{\cal{V}}_v= \frac{1}{2\p i} \int_{e_1^-}^{e_1^+} 
\frac{d\zeta}{\zeta} 
\hat{\vartheta}'_0
=\frac{1}{\p}\int_{1}^u \frac{dw}{\sqrt{1-w^2}}\hat{\vartheta}'_0
\nonumber\\
&&{\cal{V}}_1= \frac{1}{2\p i} \int_{e_0^-}^{e_0^+} 
\frac{d\zeta}{\zeta} \hat{\vartheta}'_1
=\frac{1}{\p}\int_{1}^{-\frac{1}{4\e}}\frac{dw}{\sqrt{1-w^2}}
\hat{\vartheta}'_1 \nonumber\\
&&{\cal{V}}_2= \frac{1}{2\p i} \int_{e_0^-}^{e_0^+} 
\frac{d\zeta}{\zeta} \hat{\vartheta}'_2
=\frac{1}{\p}\int_{1}^{-\frac{1}{4\e}}\frac{dw}{\sqrt{1-w^2}}
\hat{\vartheta}'_2
\label{CYperiods}
\eea
having defined the new variable
\beq
w=\frac{1}{2}\left(\zeta+\frac{1}{\zeta}\right)
\eeq
The periods in (\ref{CYperiods}) have an intersection matrix which is not in
canonical form.
In order to make contact with special K\"{a}hler geometry we turn to a canonical
basis 
\beq
v=\left(\matrix{-{\cal{V}}_v\cr 2{\cal{T}}_v+{\cal{T}}_1+2{\cal{T}}_2
-2{\cal{V}}_2+{\cal{V}}_1\cr {\cal{T}}_v+\frac{1}{2}{\cal{T}}_1+2{\cal{T}}_2
\cr {\cal{T}}_2 \cr -\frac{1}{4}{\cal{T}}_1 \cr{\cal{T}}_2+{\cal{V}}_2\cr}
\right)
\label{canonicalperiod}
\eeq
with a corresponding intersection matrix as in (\ref{canonical}).

\vspace{0.6cm}

Now we have at our disposal all the ingredients which are necessary in order 
to perform the rigid limit and compute the first gravitational corrections.
The two Calabi-Yau moduli $z=(\e,\tilde{u})$ have the following physical 
interpretation: $\e$ is the small
parameter that in the limit $\e\rightarrow 0$ leads to the decoupling of
gravity, $\tilde{u}$ is the modulus of the rigid special geometry, 
associated to the Seiberg-Witten low-energy effective action for the $SU(2)$,
$N=2$ supersymmetric Yang-Mills theory in four dimensions. Most of the work
consists in the evaluation of the integrals in (\ref{CYperiods}) to first
order in $\e$. The ones which are somewhat tricky to compute are the 
periods ${\cal{V}}_1$ and ${\cal{V}}_2$ since the integration limit
depends on $\e$. While for the other four periods it is sufficient  
first to expand the integrands in powers of $\e$ and then to perform the 
integration, this cannot be done naively for ${\cal{V}}_1$ and ${\cal{V}}_2$.
In any event a rather lengthy calculation leads to the following result
\bea
&& v_1= -\sqrt{2}\e^{\frac{1}{2}} a_D +\e^{\frac{3}{2}}\frac{\sqrt{2}}{6}
(\tilde{u}a_D+11 b_D)+{{o}}(\e^2)\nonumber\\
&& v_2=\frac{i}{\sqrt{2}}-\frac{1}{2}-\tilde{u}\e(\frac{13}{4\sqrt{2}}i+
\frac{19}{8})+\frac{\e^2}{\sqrt{2}}i(\frac{145}{48}\tilde{u}^2-\frac{479}{48})
+\frac{\e^2}{96}(271\tilde{u}^2-\frac{641}{2})+\sqrt{2}\e^{\frac{1}{2}} a_D
\nonumber\\
&&~~~~~~~-\e^{\frac{3}{2}}\frac{\sqrt{2}}{6}(\tilde{u}a_D+11 b_D)
+\frac{1}{\p i}
{\rm ln}\e\left[ \frac{i}{\sqrt{2}}+\frac{1}{2}+\tilde{u}\e (\frac{19}{8}-
\frac{13}{4\sqrt{2}}i)+\frac{\e^2}{\sqrt{2}}i(\frac{145}{48}\tilde{u}^2-
\frac{479}{48})\right. \nonumber\\ 
&&~~~~~~~~~~\left. -\frac{\e^2}{96}(271\tilde{u}^2-\frac{641}{2})\right] +
A+\e\tilde{u}B+\e^2[\tilde{u}^2C+D]
+{{o}}(\e^2)\nonumber\\
&&v_3=\frac{i}{2\sqrt{2}}-\frac{1}{2}-\tilde{u}\e (\frac{19}{8}
+\frac{13}{8\sqrt{2}}i)+\frac{\e^2}{2\sqrt{2}}i(\frac{145}{48}\tilde{u}^2-
\frac{479}{48})+\frac{\e^2}{96}(271\tilde{u}^2-\frac{641}{2})+
{{o}}(\e^2)\nonumber\\
&&v_4=-\frac{1}{4}-\frac{\sqrt{2}}{2} \e^{\frac{1}{2}} a-\frac{19}{16}
\tilde{u}\e+
\frac{\sqrt{2}}{12}\e^{\frac{3}{2}}
(\tilde{u}a+11b)+\frac{\e^2}{192}(271\tilde{u}^2 
-\frac{641}{2})+{{o}}(\e^2)\nonumber\\
&&v_5=-\frac{i}{4\sqrt{2}}+\frac{13i}{16\sqrt{2}}\tilde{u}\e-
\frac{\e^2}{4\sqrt{2}}i(\frac{145}{48}\tilde{u}^2-
\frac{479}{48})+{{o}}(\e^2)\nonumber\\
&&v_6=-\frac{1}{4}-\frac{19}{16}\tilde{u}\e+\frac{\e^2}{192}(271\tilde{u}^2 
-\frac{641}{2})-\frac{\sqrt{2}}{2} \e^{\frac{1}{2}} a_D+\e^{\frac{3}{2}}
\frac{\sqrt{2}}{12}(\tilde{u} a_D+11b_D) \nonumber\\
&&~~~~~~~+\frac{{\rm ln}\e}{2\p i}
\left[ -\frac{1}{2}-\frac{19}{8}\tilde{u}\e+\frac{\e^2}{96}(271\tilde{u}^2 
-\frac{641}{2})\right] +\a+\e\tilde{u}\b+\e^2(\tilde{u}^2\g+\d)+ o(\e^2)
\label{vectorexpanded}
\eea
We have introduced the functions
\bea
&&a_D (u)=\frac{\sqrt{2}}{\p}\int_1^{\tilde{u}}dw~ 
\sqrt{\frac{\tilde{u}-w}{1-w^2}} \qquad \qquad\quad~~
a(u)=\frac{\sqrt{2}}{\p}\int_{-1}^1dw~
\sqrt{\frac{\tilde{u}-w}{1-w^2}} \nonumber\\
&&b_D (u)=\frac{\sqrt{2}}{\p}\int_1^{\tilde{u}}dw~ w
\sqrt{\frac{\tilde{u}-w}{1-w^2}} \qquad \qquad\quad
b(u)=\frac{\sqrt{2}}{\p}\int_{-1}^1 dw~w
\sqrt{\frac{\tilde{u}-w}{1-w^2}} 
\eea
where $u=\L^2 \tilde{u}$ is the variable which appears in the Seiberg-Witten
theory \cite{SW}.
In (\ref{vectorexpanded}) we have denoted with $A$, $B$, $C$, $D$, $\a$, 
$\b$, $\g$, $\d$ various integration constants.

Using the expressions in (\ref{vectorexpanded}) one easily obtains the
K\"{a}hler potential to first order in the Calabi-Yau gravitational 
corrections
\beq
{\cal{K}}=\frac{2\p |\e|}{{\rm ln}|\e|+k}\left[i(\bar{a}a_D-\bar{a}_D a)
-\frac{8}{\p}|u|^2|\e|{\rm ln}|\e|+{\cal{O}}(\e)\right]
\label{Kahlerpotexpanded}
\eeq
where $k$ is a real constant. In (\ref{Kahlerpotexpanded}) we have dropped 
two kinds of terms, the ones which do not depend on the modulus $u$ and 
the ones which are holomorphic or antiholomorphic expressions 
that can be gauged away by a K\"{a}hler transformation. Indeed these terms do 
not contribute to the K\"{a}hler metric which is physically significant and
appears in the scalar field lagrangian. 

In order to obtain the corresponding expansion for the kinetic matrix of the
vectors one has to compute
\beq
{\cal{N}}=\left(\matrix{\bar{D}_{\bar{u}}\bar{f}_1 &
\bar{D}_{\bar{\e}}\bar{f}_1 & f_1\cr
\bar{D}_{\bar{u}}\bar{f}_2 &
\bar{D}_{\bar{\e}}\bar{f}_2 & f_2\cr
\bar{D}_{\bar{u}}\bar{f}_3 &
\bar{D}_{\bar{\e}}\bar{f}_3 & f_3\cr}\right)
\left(\matrix{\bar{D}_{\bar{u}}\bar{\chi}^1 &
\bar{D}_{\bar{\e}}\bar{\chi}^1 & \chi^1\cr
\bar{D}_{\bar{u}}\bar{\chi}^2 &
\bar{D}_{\bar{\e}}\bar{\chi}^2 & \chi^2\cr
\bar{D}_{\bar{u}}\bar{\chi}^3 &
\bar{D}_{\bar{\e}}\bar{\chi}^3 & \chi^3\cr}\right)^{-1}
\label{Nmatrix}
\eeq
Thus we have to evaluate the covariant derivatives of the period vector in
(\ref{vectorexpanded}) with respect to the modulus $u$
\bea
&&D_u\chi^1=-\sqrt{2}\e^{\frac{1}{2}}a'_D+\e^{\frac{3}{2}}\frac{\sqrt{2}}{12}
(11\frac{a_D}{\L^2}+\tilde{u}a'_D+11b_D)+
o(\e^{\frac{3}{2}}) \nonumber\\
&&D_u\chi^2=\sqrt{2}\e^{\frac{1}{2}}a'_D-\frac{1}{8\p}(13\sqrt{2}+19i)
\frac{\e{\rm ln}\e}{\L^2}+\frac{A_1\e}{\L^2}+(1+i\sqrt{2}) |\e|(\bar{a}a'_D
-\bar{a}_D a')\nonumber\\
&&~~~~~~~+\frac{3(i-\sqrt{2})\e{\rm ln}\e{\rm ln}|\e|}
{8\p\L^2({\rm ln}|\e|+k)}+o(\e)\nonumber\\
&&D_u\chi^3=-\frac{\e}{\L^2}(2+i\sqrt{2})+\frac{A_2\e}{\L^2({\rm ln}|\e|+k)}
-\frac{\sqrt{2}\p|\e|}{2({\rm ln}|\e|+k)}(1+i\sqrt{2})(\bar{a}a'_D
-\bar{a}_D a')\nonumber\\
&&~~~~~~~~~+o(\e^2{\rm{ln}}\e)\nonumber\\
&&D_u f_1=-\frac{\sqrt{2}}{2}\e^{\frac{1}{2}}a'
-\frac{\e}{\L^2}+\frac{A_3\e}{\L^2({\rm ln}|\e|+k)}
-
\frac{i\p|\e|}{2({\rm ln}|\e|+k)}(\bar{a}a'_D
-\bar{a}_D a')\nonumber\\
&&~~~~~~~~~
+\e^{\frac{3}{2}}\frac{\sqrt{2}}{24}
(11\frac{a}{\L^2}+\tilde{u}a'+11b')+
o(\e^{\frac{3}{2}})\nonumber\\
&&D_u f_2=\frac{i\e}{\sqrt{2}\L^2}+\frac{A_4\e}{\L^2({\rm ln}|\e|+k)}
+\frac{\p|\e|}{2\sqrt{2}({\rm ln}|\e|+k)}(\bar{a}a'_D
-\bar{a}_D a')+{\cal{O}}(\e^2)
\nonumber\\
&&D_u f_3=-\frac{\sqrt{2}}{2}a'_D\e^{\frac{1}{2}}+A_5\frac{\e}{\L^2}+
\frac{i\e{\rm ln}\e}{\L^2}-\frac{|\e|}{2}(\bar{a}a'_D
-\bar{a}_D a')+o(\e)
\label{covu}
\eea
and to the modulus $\e$ (the rigid limit parameter) 
\bea
&&D_{\e}\chi^1=-\frac{\sqrt{2}a_D}{2\e^{\frac{1}{2}}}\left(1-
\frac{1}{{\rm ln}|\e|+k}\right)+\sqrt{2}\e^{\frac{1}{2}}(\tilde{u}a_D
+\frac{11}{4}b_D)+o(\e^{\frac{1}{2}})\nonumber\\
&&D_{\e}\chi^2=-\frac{i-\sqrt{2}}{2\p \e}\left(1-\frac{1}{2}
\frac{|\e|}{{\rm ln}|\e|+k}\right)-\frac{B_1}{\e({\rm ln}|\e|+k)}
+\frac{\sqrt{2}a_D}{2\e^{\frac{1}{2}}}\left(1-
\frac{1}{{\rm ln}|\e|+k}\right)\nonumber\\
&&~~~~~~~~~-\frac{2}{\p}(i+\sqrt{2})\tilde{u}{\rm ln}\e
+o({\rm ln}\e)
\nonumber\\
&&D_{\e}\chi^3=-\frac{\sqrt{2}(i-\sqrt{2})}{8\e({\rm ln}|\e|+k)}
-\tilde{u}(2+i\sqrt{2})+\frac{B_2 \tilde{u}}{{\rm ln}|\e|+k}\nonumber\\
&&~~~~~~~~~-\frac{\sqrt{2}\p}{4}(i+\sqrt{2})
\frac{\bar{a}a_D-\bar{a}_D a}{{\rm ln}|\e|+k}
\left(\frac{\bar{\e}}{\e}\right)^{\frac{1}{2}}+o(\frac{1}{{\rm ln}|\e|})
\nonumber\\
&&D_{\e} f_1=\frac{1}{8\e({\rm ln}|\e|+k)}
-\frac{\sqrt{2}a}{4\e^{\frac{1}{2}}}\left(1-
\frac{1}{{\rm ln}|\e|+k}\right)
-\tilde{u}+\frac{B_3 \tilde{u}}{{\rm ln}|\e|+k}\nonumber\\
&&~~~~~~~~~-\frac{i\p(\bar{a}a_D-\bar{a}_D a)}{4({\rm ln}|\e|+k)}
\left(\frac{\bar{\e}}{\e}\right)^{\frac{1}{2}}+o(\frac{1}{{\rm ln}\e})
\nonumber\\
&&D_{\e} f_2=\frac{i}{8\sqrt{2}\e({\rm ln}|\e|+k)}
+\frac{i}{\sqrt{2}}\tilde{u}+\frac{B_4 \tilde{u}}{{\rm ln}|\e|+k}\nonumber\\
&&~~~~~~~~~~~~~+
\frac{\p(\bar{a}a_D-\bar{a}_D a)}{4\sqrt{2}({\rm ln}|\e|+k)}
\left(\frac{\bar{\e}}{\e}\right)^{\frac{1}{2}}+o(\frac{1}{{\rm ln}\e})
\nonumber\\
&&D_{\e} f_3=-\frac{i}{8\p\e}\left(\frac{{\rm ln}\e+B_5}{{\rm ln}|\e|+k}
\right)
-\frac{\sqrt{2}a_D}{4\e^{\frac{1}{2}}}\left(1-
\frac{1}{{\rm ln}|\e|+k}\right)
+B_6\tilde{u}+\frac{i}{\p} \tilde{u}{\rm ln}\e\nonumber\\
&&~~~~~~~~~~~~~~-\frac{1}{4}
(\bar{a}a_D-\bar{a}_D a)
\left(\frac{\bar{\e}}{\e}\right)^{\frac{1}{2}}+o(1)
\label{cove}
\eea
In (\ref{covu}) and (\ref{cove}) the $A_i$'s, $B_i$'s are c-numbers
expressible in terms of the constants $A,\dots,D$, and $\a,\dots,\d$ 
introduced in (\ref{vectorexpanded}).

Finally assembling all the terms and performing the matrix multiplication in
(\ref{Nmatrix}) we obtain
\bea
&&{\cal{N}}_{11}=\frac{\bar{a}'}{2\bar{a}'_D}+\frac{1}{3{\rm ln}\e+C_1}
\left(-\frac{i\p}{2}+\frac{\p(2+i\sqrt{2})}{\bar{a}'_D}\e^{\frac{1}{2}}
(\bar{a}'a_D-\bar{a}'_D a)+C_2\bar{\e}^{\frac{1}{2}}\right)+
o(\frac{\e^{\frac{1}{2}}}{{\rm ln}\e})
\nonumber\\
&&{\cal{N}}_{22}=-\frac{i\p}{2(3{\rm ln}\e+C_1)}
+o(\frac{\e^{\frac{1}{2}}}{{\rm ln}\e})
\nonumber\\
&&{\cal{N}}_{33}=\frac{\sqrt{2}{\rm ln}^2\e}{2\p(3{\rm ln}\e+C_1)}
+\frac{1}{3{\rm ln}\e+C_1}+C_3
+o(\frac{\e^{\frac{1}{2}}}{{\rm ln}\e})
\nonumber\\
&&{\cal{N}}_{23}={\cal{N}}_{32}=-\frac{1}{6}+\frac{C_4}{3{\rm ln}\e+C_1}
+o(\frac{\e^{\frac{1}{2}}}{{\rm ln}\e})
\nonumber\\
&&{\cal{N}}_{13}={\cal{N}}_{31}=\frac{1}{3}+\frac{C_5}{3{\rm ln}\e+C_1}
+\frac{1}{a'_D(3{\rm ln}\e+C_1)}\left[ \bar{\e}^{\frac{1}{2}}
\frac{{\rm ln}|\e|}{\bar{\L}^2}C_6\right. \nonumber\\
&&~~~~~~~~~~~~~~~~\left.-\e^{\frac{1}{2}}(\bar{a}'a_D-\bar{a}'_D a)
\left({\rm ln}\e(i+\sqrt{2})+C_7\right)\right]
+o(\frac{\e^{\frac{1}{2}}}{{\rm ln}\e})
\nonumber\\
&&{\cal{N}}_{12}={\cal{N}}_{21}=-\frac{i\p}{3{\rm ln}\e+C_1}
+\frac{1}{a'_D(3{\rm ln}\e+C_1)}\left[ 
\frac{2i+\sqrt{2}}{\bar{\L}^2}({\rm ln}|\e|+k-1)
+\bar{\e}^{\frac{1}{2}}\right.\nonumber\\
&&~~~~~~~~~~~~~~~~\left.+
\frac{(2+i\sqrt{2})\p}{2}\e^{\frac{1}{2}}(\bar{a}'a_D-\bar{a}'_D a)
\right]
+o(\frac{\e^{\frac{1}{2}}}{{\rm ln}\e})
\label{kineticvectors}
\eea
where the $C_i$'s are constant quantities.
The outcome of our work is collected in 
(\ref{Kahlerpotexpanded}) and
in (\ref{kineticvectors}): now we want to interpret the result, comparing it
with the expected rigid limit situation and making comments on the local,
gravitational corrections we have obtained.

\vspace{0.8cm}

The decoupling of gravity is obtained performing the $\e\rightarrow 0$ limit.
As far as the scalar field lagrangian is concerned, we have that the local 
K\"{a}hler potential in (\ref{Kahlerpotexpanded}) reduces 
to the rigid limit result found in ref. \cite{7authors}, i.e. it reproduces,
up to a normalization factor, the potential of the $SU(2)$ global 
supersymmetric theory \cite{SW}. In (\ref{Kahlerpotexpanded})  we have in 
addition the first order gravitational corrections.

Now we concentrate on the vector 
field lagrangian given in (\ref{actionvectors}): this analysis has not been
addressed in previous works. 
In order to identify the rigid limt result 
we consider the various elements of the matrix ${\cal{N}}$. From 
(\ref{kineticvectors}) we have that
\bea
&&\lim_{\e\rightarrow 0}{\cal{N}}_{11} =\frac{\bar{a}'}{2\bar{a}'_D}
\nonumber\\
&&~~~~~~\nonumber\\
&&\lim_{\e\rightarrow 0}{\cal{N}}_{22}=\lim_{\e\rightarrow 0}{\cal{N}}_{12}
=0\nonumber\\
&&~~~~~\nonumber\\
&&\lim_{\e\rightarrow 0}{\cal{N}}_{33}=
\lim_{\e\rightarrow 0}{\cal{N}}_{13}=
\lim_{\e\rightarrow 0}{\cal{N}}_{23}= {\rm{real~ constant}}
\label{rigidvectors}
\eea
Examining the terms in (\ref{actionvectors}) we learn that 
${\rm{Re}} {\cal{N}}_{IJ}$ only 
contributes to the topological part of the action. Thus the
{\em{real constant}} terms give rise to a total divergence that drops out
upon integration in the action. At this stage it is immediate to realize 
that, up to a symplectic transformation, the result in (\ref{rigidvectors})
exactly reproduces
the Seiberg-Witten $SU(2)$ effective lagrangian. It is worth to emphasize
that the above mentioned overall $\e$-dependent renormalization of the 
K\"{a}hler potential leads to a renormalization of $g_{\a\bar{\b}}$ but does not
affect ${\cal{N}}_{IJ}$: indeed the matrix ${\cal{N}}$ is of cohomological
origin and does not depend on the metric of the manifold.

The $\e$-dependent corrections in (\ref{Kahlerpotexpanded}) and 
(\ref{kineticvectors}) correspond to the first order gravitational effects.
It would be interesting to go a step further toward two distinct
directions: firstly one could consider the $N=2$ 
supergravity-Yang-Mills model resulting  from this {\em first order 
construction} and see
how it compares to the general class of supergravity actions presented in
ref. \cite{torinesi}.
Secondly one could analyze the theory obtained substituting 
(\ref{Kahlerpotexpanded}) and (\ref{kineticvectors}) in the lagrangian
$L_0+L_1$ given in (\ref{actionscalars}) and (\ref{actionvectors}). One could
interpret the
$\e$-dependent contributions as rigid supersymmetry breaking 
terms and  study their physical meaning and the type of breaking they
introduce in the theory. These and related issues are currently under
consideration.

\medskip
\section*{Acknowledgments.}

\noindent 
We thank Antoine Van Proeyen for many useful suggestions and discussions.
This work was
supported by the European Commission TMR program
ERBFMRX-CT96-0045, in which S. C. and D. Z. are associated 
to the University of Torino. 
\newpage

%\appendix

\end{document}